\documentclass{article}
\usepackage{amsmath, amssymb, amstext}
\def\1ad{\mbox{\normalsize $^1$}}
\def\2ad{\mbox{\normalsize $^2$}}
\def\3ad{\mbox{\normalsize $^3$}}
\def\4ad{\mbox{\normalsize $^4$}}
\def\5ad{\mbox{\normalsize $^5$}}
\def\6ad{\mbox{\normalsize $^6$}}
\def\7ad{\mbox{\normalsize $^7$}}
\def\8ad{\mbox{\normalsize $^8$}}
\def\makefront{
\vspace*{1cm}\begin{center}
\def\sp{
\renewcommand{\thefootnote}{\fnsymbol{footnote}}
\footnote[4]{corresponding author : \email_speaker}
\renewcommand{\thefootnote}{\arabic{footnote}}
}
\def\newtitleline{\\ \vskip 5pt}
{\Large\bf\titleline}\\
\vskip 1truecm
{\large\bf\authors}\\
\vskip 5truemm
\addresses
\end{center}
\vskip 1truecm
{\bf Abstract:}
\abstracttext
\vskip 1truecm
}
\def\({\left(}
\def\){\right)}

\newcommand{\Exc}[1]{{${\rm E}_{{#1}({#1})}$}}

\def\mxth{\mathsurround=0pt }
\def\xversim#1#2{\lower2.pt\vbox{\baselineskip0pt \lineskip-.5pt
x  \ialign{$\mxth#1\hfil##\hfil$\crcr#2\crcr\sim\crcr}}}

\renewcommand{\a}{\alpha}
\renewcommand{\b}{\beta}

\renewcommand{\d}{\delta}
\newcommand{\pa}{\partial}
\newcommand{\g}{\gamma}

\newcommand{\m}{\mu}
\newcommand{\n}{\nu}

\newcommand{\nn}{\nonumber}

\def\be{\begin{equation}}
\def\ee{\end{equation}}
\def\bea{\begin{eqnarray}}
\def\eea{\end{eqnarray}}

\newcommand{\ft}[2]{{\textstyle\frac{#1}{#2}}}

\newcommand{\eqn}[1]{(\ref{#1})}

\begin{document}
\def\titleline{Gauged maximal supergravities and\\
hierarchies of nonabelian vector-tensor systems
\renewcommand{\thefootnote}{\fnsymbol{footnote}}
\footnote{To be published in the proceedings of the 37-th International
Symposium Ahrenshoop on the Theory of Elementary Particles,
Wernsdorf, Germany, 23-27 August, 2004}
}
\def\email_speaker{{ B.deWit@phys.uu.nl}}
\def\authors{Bernard de Wit\1ad
and Henning Samtleben\2ad} 
\def\addresses{
\1ad{Yukawa Institute, Kyoto University, Kyoto, Japan {\em and} } \\
{Institute for Theoretical Physics \& Spinoza Institute,\\
  Utrecht University, The Netherlands}\\
\2ad{II. Institut f\"ur Theoretische Physik der Universit\"at Hamburg,
Hamburg, Germany}
}
\def\abstracttext{
We describe generalizations of the manifestly \Exc6 covariant
formulation of five-dimensional gauged maximal supergravity with
regard to the structure of the vector and tensor fields. We indicate
how the group-theoretical structures that we discover seem to play a
role in gauged supergravities in various space-time dimensions.}
\large
\makefront
\section{Introduction}
As is well known, gaugings are the only known supersymmetric
deformations of maximal supergravity. 
A gauging is obtained by coupling the abelian vector fields, which
arise in toroidally compactified eleven-dimensional or IIB
supergravity, to charges assigned to the elementary fields. The
resulting gauge group must be a subgroup of the duality group ${\rm
G}$ and is encoded in these charges; supersymmetry severely restricts
the possible gauge groups.  We have developed a general
group-theoretical method for determining which gaugings are
consistent \cite{dWST12}. It is based on the so-called embedding
tensor which defines how the gauge field charges are embedded into
the duality group ${\rm G}$. Treating the embedding tensor as a
spurionic object that transforms covariantly under ${\rm G}$, the
Lagrangian and transformation rules remain formally ${\rm
G}$-covariant. The embedding tensor can then be characterized
group-theoretically. When freezing it to a constant, the ${\rm
G}$-invariance is broken.  It turns out that admissible embedding
tensors must be subject to a quadratic and a linear group-theoretical
constraint. The quadratic constraint ensures that one is dealing with
a proper subgroup of ${\rm G}$ and the linear one is required by the
supersymmetry of the action upon switching on the gauge
interactions. The linear constraints are shown in
table~\ref{T-tensor-repr} for dimensions $d=3,4,5,6,7$.

However, there is a subtle issue related to the field
configuration. Because antisymmetric tensor fields of rank $p$ are
dual to antisymmetric tensors of rank $d-p-2$ in the usual setting of
a quadratic Lagrangian, the field configuration is not unique. It is
well known that this feature is of relevance for certain gaugings. For
instance, in the standard framework only the gauge fields belonging to
the adjoint representation of the gauge group can carry the
corresponding charges. Problems with charged vector fields in other
representations are usually circumvented by dualizing these fields, 
after which the gauging can still proceed
\cite{TownPilcNieuw,PPvN,GunaRomansWarner}. However, the fact that
the field representation must first be carefully adapted before
switching on the gauge coupling, hampers a general analysis of the
gaugings. Moreover, the manifest ${\rm G}$-covariance is affected by the fact
that the fields will no longer constitute ${\rm G}$ representations. In
\cite{dWST5} a novel system of vector-tensor fields was adopted for
five-dimensional supergravity, in
which the tensor and the vector fields each constitute a complete
${\rm G}$ representation. Because of the presence of additional gauge
transformations, also depending on the embedding tensor, the number of
degrees of freedom remains, however, unchanged. This extended field
configuration can thus accommodate any possible gauging and the
group-theoretical analysis proceeds in a uniform way. The
viability of this formulation can be established a posteriori by
rederiving all the supersymmetry transformations and the Lagrangian in
the new formulation;
indeed, apart from the two types of group-theoretical constraints on
the embedding tensor, no other conditions are necessary. 

Motivated by this result we have analyzed whether similar results can also
be derived for other space-time dimensions. Here we report on the 
outcome of this analysis, which indicates that hierarchies of
vector-tensor systems do appear in other dimensions, in a similar
group-theoretical setting. We first review the results of \cite{dWST5}
relevant for the vector-tensor couplings and their
generalization. Finally we discuss their application in various
dimensions.  

\begin{table}
\begin{center}
\begin{tabular}{l l  l l  }\hline
~&~&~&~\\[-4mm]
$d$ &${\rm G}$& ${\rm H}$ & $\Theta$  \\   \hline
~&~&~&~\\[-4mm]
7   & ${\rm SL}(5)$ & ${\rm USp}(4)$  & ${\bf 10}\times {\bf 24}= {\bf
  10}+\underline{\bf 15}+  \underline{\bf 40}+ {\bf 175}$  \\[1mm]
6  & ${\rm SO}(5,5)$ & ${\rm USp}(4) \times {\rm USp}(4)$ & 
  ${\bf 16}\times{\bf 45} =
  {\bf 16}+ \underline{\bf 144} + {\bf 560}$ \\[.8mm]
5   & ${\rm E}_{6(6)}$ & ${\rm USp}(8)$ & ${\bf 27}\times{\bf 78} = 
  {\bf 27} + \underline{\bf 351} + {\bf 1728}$  \\[.5mm]
4   & ${\rm E}_{7(7)}$ & ${\rm SU}(8)$  & ${\bf 56}\times{\bf 133} = 
  {\bf 56} + \underline{\bf 912} + {\bf 6480}$   \\[.5mm]
3   & ${\rm E}_{8(8)}$ & ${\rm SO}(16)$ & ${\bf 248}\times{\bf 248} = 
  \underline{\bf 1} + {\bf 248} + \underline{\bf 3875} +{\bf 27000} 
  +  {\bf 30380}$ 
\\ \hline
\end{tabular}
\end{center}
\caption{\small
Decomposition of the embedding tensor $\Theta$ for maximal
supergravities in various space-time dimensions in terms of irreducible
${\rm G}$ representations.  Only the underlined representations are
allowed according to the representation constraint. The R-symmetry
group ${\rm H}$ is the maximal compact subgroup of ${\rm G}$.
}\label{T-tensor-repr}
\end{table}

\section{Some results for $d=5$ with a slight generalization}
\label{embed}
In maximal five-dimensional supergravity, the (abelian) vector fields
$A_\m^M$ transform in a representation  
$\overline{\bf 27}$ of ${\rm G}={\rm E}_{6(6)}$:  $\d A_\m{}^M =
-\Lambda^\a (t_\a)_N{}^M\,A_\m{}^N$,  where the 78 independent \Exc6
generators are denoted by  
$(t_\a)_M{}^N$. Because the gauge group is a subgroup of \Exc6, its
generators $X_M$ are decomposable in terms of the $t_\a$, {\it  i.e.},   
\be
\label{X-theta-t}
X_M = \Theta_M{}^\a\,t_\a\;,
\ee
where $\a=1,2,\ldots,78$ and $M=1,2,\ldots,27$. 
The gauging is thus encoded in a real {\it embedding tensor}
$\Theta_{M}{}^{\alpha}$ assigned to the ${\bf 27}\times{\bf 78}$
representation of \Exc6. The embedding tensor acts as
a projector whose rank equals the dimension of the gauge group (up to
abelian gauge fields corresponding to a possible central extension of
the gauge algebra). 
The $X_M$ generate a group and thus define a Lie algebra, ${[X_M,X_N]}
= f_{MN}{}^P\,X_P$, 
with $f_{MN}{}^P$ the as yet unknown structure constants of the gauge
group. Hence the embedding tensor must satisfy the closure condition,
\be
\label{gauge-gen}
\Theta_M{}^\a\,\Theta_N{}^\b \,f_{\a\b}{}^{\g}= f_{MN}{}^P\,
\Theta_P{}^\g\,,
\ee
were $f_{\a\b}{}^\g$ denotes the structure constants of \Exc6, 
according to $[t_\a,t_\b]= f_{\a\b}{}^\g\,t_\g$. Consequently the
structure constants $f_{MN}{}^P$ satisfy the 
Jacobi identities, but only in the subspace projected by the embedding tensor!

Once the gauge group is specified, covariant derivatives are
introduced, $D_\m = \partial_\m - g\,A_\m{}^M\,X_M$, where $g$ denotes
the gauge coupling constant. They lead to covariant field strengths, 
\be
\label{eq:cov-FS} 
\Theta_M{}^\a \,{\cal F}_{\m\n}^M = \Theta_M{}^\a (\pa_\m A_\n{}^M  - \pa_\n
A_\m{}^M   -g\,f_{NP}{}^M \, A_\m{}^N\, A_\n{}^P)\,.  
\ee
Gauge field transformations are given by
\be
\label{eq:gauge-transf}
\Theta_M{}^\a \,\d A_\m{}^M = \Theta_M{}^\a\, (\pa_\mu \Lambda^M - g\,
f_{NP}{}^M \,A_\m{}^N\,\Lambda^P)\,.
\ee
Because of the contraction with the embedding tensor, the
above results apply to only a subset of the gauge fields; the
remaining ones do not appear in the covariant derivatives and are not
directly involved in the gauging. The gauge fields that do
appear in the covariant derivatives, are only determined up to
additive contributions by gauge fields that vanish upon contraction
with $\Theta_M{}^\a$. 

While the gauge fields involved in the gauging should transform in the adjoint
representation of the gauge group, the gauge field charges should also
coincide with $X_M$ in the $\overline{\bf 27}$ representation. Therefore
$X_{MN}{}^P\equiv(X_M)_N{}^P$  must decompose into the adjoint
representation of the gauge  group plus possible extra terms which
vanish upon contraction with the embedding tensor, 
\be
\label{adjoint}
X_{MN}{}^P \,\Theta_P{}^\a \equiv \Theta_M{}^\b \,t_{\b N}{}^P\,
 \Theta_P{}^\a  = - f_{MN}{}^P \,\Theta_P{}^\a \,.
\ee
Note that \eqn{adjoint} is the analogue of \eqn{gauge-gen} in the
$\overline{\bf 27}$ representation.  
The combined conditions \eqn{gauge-gen} and \eqn{adjoint} imply that $\Theta$
is invariant under the gauge group and yield the \Exc6-covariant condition,
which can be conveniently written as,
\be
\label{eq:X-closure}
{[X_M,X_N]} = -X_{MN}{}^P\,X_P\,.
\ee
Note that $X_{MN}{}^P$ is not assumed to be antisymmetric in $M$ and $N$,
but only the antisymmetric component contributes upon contraction with
the embedding tensor. One can prove that
(\ref{eq:X-closure}) suppresses the $\overline{\bf 27}+{\bf 1728}$
representation in products quadratic in the embedding tensor
components. To prove this, the linear constraint on the embedding
tensor is relevant. As shown in table~\ref{T-tensor-repr} the
embedding tensor must belong to the ${\bf 351}$ representation.

The tensor $X_{MN}{}^P$ transforms in the ${\bf 351}$ representation
of \Exc6, just as the embedding tensor itself. Exploiting the
symmetric \Exc6-invariant tensors $d_{MNP}$ and $d^{MNP}$, one
constructs an antisymmetric tensor  
$Z^{MN}\equiv X_{PQ}{}^M\, d^{NPQ}$, which thus also belongs to the
${\bf 351}$ representation. For what follows the inverse
equation is relevant, 
\begin{equation}
\label{X-dZ}
X_{(MN)}{}^P = d_{MNQ}\,Z^{PQ}  \,.
\end{equation}
According to \cite{dWST5} one can derive the following relations, 
\begin{equation}
\label{eq:Z-X-orthogonal}
Z^{MN}\,\Theta_N{}^\a = 0\,,\qquad Z^{MN}\,X_{N}= 0\,,\qquad
X_{MN}{}^{[P}\,Z^{Q]N}= 0\,,
\end{equation}
where, in the second equation, $X_M$ is taken in an arbitrary representation. 
The third equation implies that $Z^{MN}$ is {\it invariant} under the
gauge group.  

Rather than continuing this presentation of $d=5$ results, we consider 
a slight generalization which will contain the $d=5$ 
results as a special case. Namely, we replace $Z^{MN}$ by a
tensor $Z^{M,I}$ belonging to a product representation of ${\rm G}$,
where ${\rm G}$ will be kept arbitrary. 
Here indices $M, N,\ldots$ remain, as before, related to the representation
to which the gauge fields have been assigned, but the indices $I,J,\ldots$
belong to some other irreducible representation of ${\rm G}$. The new
tensor $Z^{M,I}$ now appears in a modification of 
(\ref{X-dZ}), 
\begin{equation}
\label{X-dZ-new}
X_{(MN)}{}^P = d_{I,MN}\,Z^{P,I}  \,,
\end{equation}
where $d_{I,MN}$ is a ${\rm G}$-invariant tensor, {\it i.e.},
$t_{\alpha}{}_{M}{}^{P}\,d_{I,PN}+t_{\alpha}{}_{N}{}^{P}\,d_{I,PM}+ 
t_{\alpha}{}_{I}{}^{J}\,d_{J,MN} =0$, which is symmetric in $M$ and $N$.
Obviously (\ref{eq:X-closure}) will now apply to the two
representations separately, with generators $X_{MN}{}^P$ and
$X_{MI}{}^J$, respectively. It then 
follows that $X_{MN}{}^P$, $X_{MI}{}^J$ and $Z^{M,I}$ are gauge
invariant tensors which take their values in representations belonging
to the embedding tensor. Note, however, that we are not insisting that
the embedding tensor belongs to an irreducible
representation. Furthermore it follows that
\begin{equation}
\label{eq:Z-X-orth-new}
Z^{M,I}\,X_{MN}{}^P = 0= Z^{M,I}\,X_{MJ}{}^K\,, 
\end{equation}
We expect that (\ref{eq:Z-X-orth-new}) and (\ref{eq:X-closure}) are
equivalent versions of the quadratic constraint (of course, upon the
condition that the linear constraint on the embedding tensor
holds). 

It then follows that the tensor $Y_{IM}{}^J$, defined by 
\begin{equation}
\label{eq:Y-def}
Y_{IM}{}^J \equiv X_{MI}{}^J + 2\,d_{I,MN}\,Z^{N,J}\,,
\end{equation}
satisfies an orthogonality equation 
\begin{equation}
\label{eq:Z-Y-orthogonal}
Z^{M,I}\,Y_{I N}{}^J = 0\,. 
\end{equation}
Furthermore one can derive the following property,
\begin{equation}
\label{Jacobi-X}
 {X_{[MN]}{}^P\, X_{[QP]}{}^R + X_{[QM]}{}^P\, X_{[NP]}{}^R  + X_{[NQ]}{}^P \,
 X_{[MP]}{}^R} = - Z^{R,I}\,d_{I,P[Q}\, X_{MN]}{}^P \,.  
\end{equation}
This equation is relevant in the next section where we will employ
$X_{[MN]}{}^P$ as an extension of the gauge group structure constants
$f_{MN}{}^P$, which satisfies the Jacobi identity up to terms
proportional to $Z$. As the right-hand side vanishes upon contraction
with the embedding tensor $\Theta_R{}^\a$, we see that the
$X_{[MN]}{}^P$ satisfy the Jacobi identity in the subspace projected
by the embedding tensor, just as the gauge group structure constants. 

The above results are directly compatible with the results for
$d=5$ by removing the distinction between the indices $M,\ldots$ and
$I,\dots$. Note that $Y_{IM}{}^J$ becomes equal to
$-X_{IM}{}^J$, so that (\ref{X-dZ-new}) and
(\ref{eq:Z-Y-orthogonal}) will coincide.  
\section{Nonabelian vector-tensor systems}
\label{vector-tensor}
In the previous section we presented the transformations
\eqn{eq:gauge-transf} and the field strengths \eqn{eq:cov-FS} of the
vector fields, which were the conventional ones 
except that these expressions were contracted with the embedding
tensor and thus apply to only part of the fields. One way to make them
exact is to remove the fields that are projected to zero by the
embedding tensor, for instance, by dualizing them into tensor
fields. Because this would affect the duality invariance, we will set
up an alternative formulation where vectors and tensor gauge fields
constitute full representations of the duality group; at the same
time, we will introduce extra gauge invariances so that the total
number of degrees of freedom remains the same, irrespective of the
embedding tensor that has been adopted.  

To see how this works let us follow \cite{dWST5} and  consider the
gauge transformations of the vector fields,
\begin{equation}
  \label{eq:A-var}
\delta A_\mu{}^M = \partial_\mu\Lambda^M - g\,
X_{[PQ]}{}^M\,\Lambda^P\,A_\mu{}^Q - g\,Z^{M,I}\,\Xi_{\m\,I}\,,  
\end{equation}
where $\Lambda^M$ is the usual gauge transformation parameter and the
transformations proportional to  $\Xi_{\m\,I}$ allow us to gauge away
those vector 
fields that are perpendicular to the embedding tensor. This equation
is thus an extension of (\ref{eq:gauge-transf}). The usual field
strength is not covariant and we define a modified field strength,
\be 
\label{eq:modified-fs}
{\cal H}_{\m\n}{}^M= 
{\cal F}_{\mu\nu}{}^M  
+ g\, Z^{M,I} \,B_{\m\n\,I}\;,
\ee
which contains tensor fields $B_{\m\n\,I}$.
Here we use the notation, 
\be
{\cal  F}_{\mu\nu}{}^M =\pa_\m A_\n{}^M -\pa_\n A_\m{}^M + g\,
X_{[NP]}{}^M \,A_\m{}^N A_\n{}^P \,. 
\ee
By making a suitable choice for the
transformation rule of the tensor field $B_{\mu\nu I}$, 
\bea
\label{eq:dB}
\delta B_{\m\n\,I} &=&{} 
2\,\pa_{[\m} \Xi_{\n]I} - g\,X_{MI}{}^J \,A_{[\m}{}^M\,\Xi_{\n]J} 
+ g\,\Lambda^M\,X_{MI}{}^J \,B_{\mu\nu\,J}   \nn\\[.5ex]
&&
{}+\Lambda^M\Big[- d_{I,MN} \,{\cal F}_{\mu\nu}{}^N
+ g\,d_{I,N[M}\,X_{PQ]}{}^N \,A_\mu{}^P A_{\nu}{}^Q\Big] 
- g\, Y_{IM}{}^J \, \Phi_{\mu\nu\,J}{}^M \,,
\eea
the modified field strength  transforms covariantly, 
$\d{\cal H}_{\m\n}{}^M  = - g\, X_{PN}{}^M \,\Lambda^P\,{\cal
H}_{\m\n}{}^N$. 
Here we introduced a new tensor gauge transformation with parameter
$\Phi_{\mu\nu\,J}{}^M$, which does not interfere with the field
strength (\ref{eq:modified-fs}) in view of
(\ref{eq:Z-Y-orthogonal}).\footnote{
  Observe that the decomposition of (\ref{eq:dB}) is ambiguous, as we
  can  redefine the parameter $\Phi_{\mu\nu\,I}{}^M$ with terms
  proportional to $\Xi_{\mu\,I}$ and $\Lambda^M$. }  
In $d=5$ dimensions this option is not relevant, as the tensor fields
appeared only in the combination $Z^{M,I}\,B_{\mu\nu\,I}$. 

In the general case we are about to discover the need for yet another
tensor field $S_{\mu\nu\rho\,I}{}^M$, and in this way we should expect
to generate a whole hierarchy of tensor fields. Before continuing, let
us make a number of comments. First of all, it is 
important to note that the covariant derivative does not transform
under tensor gauge  transformations, by virtue of
\eqn{eq:Z-X-orthogonal}, and also the Ricci identity, 
${[}D_\m,D_\n] = -g\,{\cal F}_{\m\n}{}^M \,X_M$, will still hold.
Secondly, to verify the consistency of the transformations obtained
so far, one may consider the commutator algebra. While the tensor
transformations commute,  
\be
{[}\d({\Xi_1}), \d({\Xi_2})] = {[}\d({\Phi_1}), \d({\Phi_2})] =
{[}\d({\Phi}), \d({\Xi})]=  0\,,
\ee 
the commutator of a vector and a tensor gauge transformation gives rise to
tensor gauge transformations,
\be
{[}\d({\Lambda}), \d({\Xi})] = \d(\tilde\Xi)+\d(\tilde\Phi) \,,
\ee
with
\bea
\tilde\Xi_{\m\,I} &=& {}- \ft12 g\,X_{MI}{}^J\,\Xi_{\m\,J}\,\Lambda^M
\;, \nn\\[.5ex] 
\tilde{\Phi}_{\mu\nu\, I}{}^{M} &=&
( \partial_{[\mu}\Xi_{\nu]I}
-\ft13g\, X_{NI}{}^{J}\,A_{[\mu}{}^{N}\,\Xi_{\nu]J})\Lambda^{M}
-\ft13g\,X_{[NP]}{}^{M}\,\Lambda^N A_{[\mu}{}^{P}\,\Xi_{\nu]I}
+\ft13 A_{[\mu}{}^M\,   \tilde \Xi_{\nu]I}\,. \nn\\
&&{~}
\eea
Likewise, 
\be
{[}\d({\Lambda}), \d({\Phi})] = \d(\tilde\Phi)\;,
\qquad
\tilde{\Phi}_{\mu\nu\, I}{}^{M} =
g\, \Lambda^P (X_{PN}{}^M \Phi_{\mu\nu\, I}{}^{N}- X_{PI}{}^J
\Phi_{\mu\nu\, J}{}^{M}) \;. 
\ee
The remaining commutator of two vector gauge transformations gives
rise to a combination of vector and tensor gauge
transformations,
\be
{[}\d({\Lambda_1}), \d({\Lambda_2})] = \d(\tilde\Lambda)+
\d(\tilde\Xi) + \d(\tilde\Phi) \;, 
\ee
where 
\bea
 \tilde \Lambda^M &=& g\, \Lambda_1^N\,\Lambda_2^P\,X_{[NP]}{}^M\,, 
 \nonumber\\
\tilde\Xi_{\m\,I} &=&
{}- g\,\Lambda_1^M\,\Lambda_2^N\,
d_{I,Q[M}\,X_{NP]}{}^Q\,A_\m{}^P \,, \nn\\
\tilde{\Phi}_{\mu\nu\, I}{}^{M}&=&
{}  \Lambda_1^{[M}\Lambda_2^{N]}\Big[-\ft23 d_{I,NP}\,{\cal
F}_{\mu\nu}{}^P 
+ g\, d_{I,R[N}\,
X_{PQ]}{}^{R}\, A_{\mu}{}^{P}A_{\nu}{}^{Q}\Big]  \;. 
\eea

We can now continue and introduce a rank-3 antisymmetric gauge
field $S_{\mu\nu\rho\,I}{}^M$. Just as before we propose a covariant
field strength, but now for the tensor field $B_{\mu\nu\,I}$, 
\bea
{\cal H}_{\mu\nu\rho\,I} &\equiv&
3\,\Big[ D_{[\mu} B_{\nu\rho]\,I} +2
\,d_{I,MN}\,A_{[\mu}{}^{M}(\pa_{\nu} A_{\rho]}{}^N+ 
\ft13 g X_{[PQ]}{}^{N}A_{\nu}{}^{P}A_{\rho]}{}^{Q})\Big] \nn\\
&&{} 
+ g\,Y_{IM}{}^J\, S_{\mu\nu\rho\,I}{}^M \,,
\eea
with the covariant derivative 
$D_{[\mu}B_{\nu\rho]I}=\partial_{[\mu} B_{\nu\rho]I}
-gX_{MI}{}^{J}\,A_{[\mu}{}^{M}\,B_{\nu\rho]J}\,$.
This field strength should remain invariant under the tensor
and transform covariantly under the vector gauge
transformations, {\it i.e.}, $\d{\cal H}_{\mu\nu\rho\,I} =
g\, \Lambda^{M}\,X_{MI}{}^{J}\,{\cal H}_{\mu\nu\rho\,J}$. 
This can be achieved provided we assign the following transformations
to $S_{\mu\nu\rho\,I}{}^M$, 
\bea
\d\,S_{\mu\nu\rho\,I}{}^{M}
&\!=\!&
 g\,\Lambda^N X_{NI}{}^J\, S_{\mu\nu\rho J}{}^M  
- g\,\Lambda^N X_{NP}{}^M\, S_{\mu\nu\rho I}{}^P \nonumber\\
&&{} +3\,D_{[\mu}\Phi_{\nu\rho]I}{}^{M} 
+3\,A_{[\mu}{}^{M}\,D_{\nu}\Xi_{\rho]I}
+3 \,\partial_{[\mu}A_{\nu}{}^{M} \,\Xi_{\rho]\,I}
-2 g \, d_{I,NP}\,Z^{P,J}
 A_{[\mu}{}^{M}A_{\nu}{}^{N}\Xi_{\rho]J}
 \nonumber \\[.5ex]
&&{}
+4\, d_{I,NP}\,\Lambda^{[M}\,A_{[\mu}{}^{N]}\, \partial_{\nu}A_{\rho]}{}^{P}
+2 g\,X_{NI}{}^{J}\,d_{J,PQ} \,
\Lambda^{Q}\,A_{[\mu}{}^{M}A_{\nu}{}^{N}A_{\rho]}{}^{P}   
\,.
\eea
This procedure can in principle be continued to higher-rank tensor
fields. Note, for instance, the additional orthogonality relation,
$Y_{IM}{}^J \,d_{J,(NP}\,\delta_{Q)}{}^M =0$.

\section{Vector-tensor couplings in various dimensions}
So far, our arguments have been rather abstract without giving any
indication as to why the vector-tensor hierarchy that we exhibited
should have a role to play in gauged maximal supergravities (apart from
the five-dimensional theory, which inspired this analysis). In this
last section we will briefly discuss this issue. As it turns out, the
group-theoretical setting for generating the vector-tensor hierarchies
leads to results that fit in very naturally with what is known about
the ungauged theories and with what one expects for the gauged
versions. The tensor representations tend to come out such that
tensors that are dual to each others transform also in representations
of ${\rm G}$ that are mutually conjugate. Furthermore, the embedding tensors 
always seem to enable the covariant construction of the relevant topological
couplings. There remain open questions, in particular for even
space-time dimensions. In any case, our conclusion is that the
group-theoretical structure of the vector-tensor hierarchies is such that
they clearly play a role in gauged supergravities. The scope of this
paper does not allow us to present the couplings in any detail but
we hope to report on this elsewhere. We close with a brief perusal of
gauged maximal supergravities in various dimensions. \\[.2ex]

\noindent{\it $d=4$ supergravity}~: Here the duality group is ${\rm
G}={\rm E}_{7(7)}$ and the  
vector fields transform in the ${\bf 56}$ representation. An obvious
subtlety (just as in $d=6$ dimensions) is, that the 
duality group applies to the equations of motion and not to the
Lagrangian, whereas the ${\bf 56}$ representation of the vector fields
covers both the electric and the magnetic potentials. We do not view
this as a negative feature, but rather we expect that eventually the
observations of this paper will have some positive impact on this
question. As shown in table~\ref{T-tensor-repr} the embedding 
tensor $\Theta_{M}{}^{\alpha}$ belongs to the ${\bf 912}$
representation of \Exc7. Consequently we know that the fully symmetric
part of $X_{MN}{}^Q\,\Omega_{PQ}$ must vanish, where $\Omega$ denotes
the invariant skew-symmetric tensor of \Exc7.\footnote{
  This is discussed in a forthcoming paper \cite{dWST4}.} 
Obviously there exists an \Exc7-invariant tensor, $d_{\alpha,MN}=
t_{\alpha\,M}{}^P\,\Omega_{NP}$. Upon multiplication by 
$\Theta_M{}^\alpha$ it follows that $X_{(MN)}{}^P = -\ft12
\Theta_Q{}^\alpha\, \Omega^{QP} \, d_{\alpha,MN}$, so that
(\ref{X-dZ-new}) holds with $Z^{M,\alpha} = \ft12
\Omega^{MN}\,\Theta_N{}^\alpha$. This implies that 
rank-2 tensor fields $B_{\mu\nu\,\alpha}$ will exist in the ${\bf 133}$
representation of \Exc7. Note that $Z^{M,\beta}\,\Theta_M{}^\alpha=0$
by virtue of the quadratic constraint on the embedding tensor. 

At this point we have no complete Lagrangian, but we note that the
embedding tensor $\Theta_M{}^\alpha$ has precisely the right structure
to construct an \Exc7-invariant topological coupling 
${\cal H}\wedge B$ of the vector field strengths ${\cal
H}_{\mu\nu}{}^{M}$ and the rank-2 tensor fields
$B_{\mu\nu\,\alpha}$.\\[.2ex]   

\noindent{\it $d=5$ supergravity}~:
We have already discussed the characteristic features of this case
(see also, \cite{dWST5}), with the vector fields 
transforming in the ${\bf \overline{27}}$ representation and the
rank-2 tensors in the ${\bf 27}$ representation of \Exc6. 
The $d$-symbol is the \Exc6-invariant symmetric tensor
$d_{M,NP}=d_{MNP}$. The tensor $Z^{MN}$ is 
antisymmetric and belongs to the ${\bf 351}$ representation of \Exc6,
just as the embedding tensor. It enables the construction of an \Exc6-invariant
topological term $B\wedge DB$, which is known to appear in the $d=5$
theories on a par with the Chern-Simons term for the vector gauge
fields. 
 
The tensor $Y$ from (\ref{eq:Y-def}) reduces to $Y_{MN}{}^{P}=-
X_{MN}{}^{P}= -\Theta_{M}{}^{\alpha}\,t_{\alpha\,N}{}^{P}$, so that rank-3
tensors appear only in the form 
$t_{\alpha\,N}{}^{P}\,S_{\mu\nu\rho\,P}{}^{N}\equiv
S_{\mu\nu\rho\,\alpha}$ and thus belong to the adjoint ${\bf 78}$
representation of \Exc6. These fields do not appear in the
Lagrangian.\\[.2ex] 

\noindent{\it $d=6$ supergravity}~: The duality group
equals ${\rm G}= {\rm E}_{5(5)}={\rm SO}(5,5)$ with vector fields
transforming in the ${\bf 16}_c$ representation of  ${\rm SO}(5,5)$.
The embedding tensor must belong to the ${\bf 144}_{c}$ 
representation. This is a vector-spinor representation so that the 
embedding tensor can be written as 
$\Theta_{\underline\alpha}{}^{[mn]}=
\theta^{\underline\beta [m}\,
\Gamma^{n]}{}_{\underline\beta\underline\alpha}$. 
Here $\Gamma_{m\,\underline\alpha\underline\beta}$ is symmetric in the
spinor indices $\underline\alpha,\underline\beta$ and corresponds to
the ${\rm SO}(5,5)$ gamma matrices restricted to the chiral subspace
(after contraction with the charge conjugation matrix). The obvious
choice for the $d$-symbol is $d_{m\,\underline\alpha\underline\beta}=
\Gamma_{m\,\underline\alpha\underline\beta}$.  The 45 
generators of the duality group in the ${\bf 16}_c$ representation 
are proportional to $\Gamma_{mn}$, so that we derive the following
relation for the gauge generators, 
\be
X_{(\underline\alpha\underline\beta)}{}^{\underline{\gamma}} =
\Theta_{(\underline\alpha}{}^{mn}\,(\Gamma_{mn})_{\underline\beta)}
{}^{\underline\gamma}
=~ \theta^{\underline\delta m}\,\Gamma^{n}
{}_{\underline\delta(\underline\alpha}\,
(\Gamma_{mn})_{\underline\beta)}{}^{\underline\gamma}
=  \theta^{\underline\gamma m}
d_{m,\underline\alpha\underline\beta}\;.
\ee
{}From (\ref{X-dZ-new}) we can then read off the tensor 
$Z^{\underline\alpha,m}= \theta^{\underline\alpha m}$. 
Hence, we expect rank-2 tensors $B_{\mu\nu\,m}$ in the ${\bf 10}$
representation, which are indeed present in the ungauged theory. 

Using the corresponding generators in the ${\bf 10}$ representation,
$(t_{mn})_{p}{}^{q}=4\eta_{p[q}\,\delta_{n]}{}^{q}$ 
one finds from (\ref{eq:Y-def}) for the tensor
$Y_{m\underline{\alpha}}{}^{n}$: 
\be
Y_{m\underline{\alpha}}{}^{n} =
2\,\theta^{\underline\delta p}\,\Gamma^{q}{}_{\underline\delta\underline\alpha}
\, \eta_{mp}\,\delta_{q}{}^{n} -
2\,\theta^{\underline\delta p}\,\Gamma^{q}{}_{\underline\delta\underline\alpha}
\,\eta_{mq}\,\delta_{p}{}^{n}
+2\, \theta^{\underline\beta n}\,
\Gamma_{m\,\underline\alpha\underline\beta}
= 2\,\theta^{\underline\delta}{}_m \,
\Gamma^{n}{}_{\underline\delta\underline\alpha} \;.
\ee
This shows that three-forms are generated in the combination, 
$\Gamma^{m}{}_{\underline\alpha\underline\beta}
\,S_{\mu\nu\rho\,m}{}^{\underline\beta}\equiv 
S_{\mu\nu\rho\,\underline\beta}\,$. 
Indeed, in six dimensions, three-form tensor fields are dual to vector fields
and thus appear in the conjugate representation. The validity of the
orthogonality relations
(\ref{eq:Z-X-orthogonal},\ref{eq:Z-Y-orthogonal}) is ensured by the
quadratic constraint on the embedding tensor ({\it c.f.} eq. (6.16) of
\cite{dWST5}). 
The embedding tensor $\theta^{\underline\alpha m}$
is of the right structure to write down a topological
coupling ${\cal H}\wedge S$ between the rank-2 tensor field strengths
${\cal H}_{\mu\nu\rho\,I}$ and the new rank-3 tensor fields
$S_{\mu\nu\rho\,\underline\alpha}$. \\[.2ex]

\noindent{\it $d=7$ supergravity}~: The duality group equals ${\rm G}= {\rm
E}_{4(4)}={\rm SL}(5)$ with the vector fields transforming in the
${\bf \overline{10}}$ representation. The embedding tensor must take 
its values in the ${\bf 15}+{\bf \overline{40}}$
representation,\footnote{
  In \cite{dWST12} we incorrectly omitted the ${\bf \overline{40}}$
  representation. Its presence was deduced in \cite{dWST5}. } 
so that it can be written as 
 $\Theta_{[mn]}{}_{p}{}^{q}=
\delta_{[m}{}^{q}\,Y^{\phantom{l}}_{n]p} -2\,\varepsilon_{mnprs}\,Z^{rs,q}$. 
where $Y^{mn}$ is a symmetric tensor and $Z^{mn,p}$ is antisymmetric
in $[m,n]$ and satisfies the irreducibility constraint $Z^{[mn,p]}=0$. 
The gauge generators in the ${\bf 10}$-representation then take the form
\be
(X_{mn}){}_{pq}{}^{rs} =
-2\,\delta^{rs}{}_{[m[p}\,Y^{\phantom{l}}_{q]n]}
-4\,\varepsilon^{\phantom{l}}_{mntu[p}\,Z^{tu,[r}\,\delta_{q]}{}^{s]}
\;,
\ee
from which we find
\be
(X_{mn}){}_{pq}{}^{rs}+(X_{pq}){}_{mn}{}^{rs} =
-4\,\varepsilon_{tmnpq}\,Z^{t[r,s]}
= 2\,Z^{rs,t}\,d_{t,[mn][pq]}\;, 
\ee
where the $d$-symbol was defined by $d_{r,[mn][pq]}=\varepsilon_{rmnpq}$.
This establishes that $Z$ is the same tensor that appears in 
(\ref{X-dZ-new}). Hence we expect tensors $B_{\mu\nu\,m}$ transforming
in the ${\bf 5}$ representation, which, indeed, are known to be present in the
ungauged theory. 

Likewise one can evaluate the combination
(\ref{eq:Y-def}) and show that the tensor $Y_{p[mn]}{}^q$ is given by
$Y_{p[mn]}{}^q = \delta^q{}_{[m}\,Y_{n]p}$.  Therefore there will be
rank-3 gauge fields appearing in the combination, 
$\delta^{q}{}_{[m}\,Y_{n]p} \,S_{\mu\nu\rho\,q}{}^{mn}= -
Y_{pm}\,S_{\mu\nu\rho\, n}{}^{mn}\equiv - 
Y_{pq}\,S_{\mu\nu\rho}{}^{q}\,$.
The rank-3 tensors thus transform in the $\overline{\bf 5}$
representation and are thus dual to the rank-2 tensors.

Finally the $Y$-component of the embedding tensor is of the right
structure to write down a topological coupling $S \wedge D S$ of these
rank-three tensor fields $S_{\mu\nu\rho}{}^{I}$. The embedding tensors
with $Z=0$ encode all the ${\rm CSO}(p,q,r)$
gaugings (where $p+q+r=5$), of which a subclass was constructed in
\cite{PPvN}. In this subclass all rank-2 tensors can be gauged away,
and one is left with only rank-3 tensor fields.

\vspace{3mm}
\noindent
We thank Mario Trigiante for valuable discussions. This work is partly
supported by EU contracts MRTN-CT-2004-005104 and
MRTN-CT-2004-503369, the INTAS contract 03-51-6346, and the DFG grant
SA 1336/1-1.  

\providecommand{\href}[2]{#2}
\begingroup\raggedright\endgroup

\end{document}